# Fast Identification of Saturated Cut-sets using Graph Search Techniques

Reetam Sen Biswas, Anamitra Pal, Trevor Werho, Vijay Vittal
School of Electrical, Computer, and Energy Engineering
Arizona State University
Tempe, Arizona-85287, USA

*Abstract*—**When multiple outages occur in rapid succession, it is important to know quickly if the power transfer capability of different interconnections (or cut-sets) of the power network are limited. The algorithm developed in this paper identifies such limited cut-sets very fast, thereby enhancing the real-time situational awareness of power system operators. The significance of the proposed approach is described using the IEEE 39-bus test system, while its computational benefits are demonstrated using relatively large test-cases containing thousands of buses. The results indicate that the proposed network analysis can estimate the impact of an outage on any cut-set of the system and screen out the cut-set that gets saturated by the largest margin, very quickly.**

## I. Introduction

When multiple outages occur in rapid succession, such as, during extreme event scenarios, it is imperative that the power system operators quickly identify the critical transmission interconnections that have limited power transfer capability. Analysis of prior blackouts have proved that lack of this knowledge may be catastrophic for the system [1], [2]. Werho et al. showed in [3] that a critical interconnection may not necessarily be a single transmission line whose status can be monitored; i.e., a critical interconnection can be a collection of multiple lines. As such, quantifying the impact of a contingency on a critical interconnection (or cut-set) of the power system is a task worth undertaking.

A variety of metrics, such as *node degree* [4], *betweenness indices* [5], [6], *risk graph* [7], *geodesic vulnerability* [8], and *modified centrality indices* [9] have been proposed previously to investigate the topology and network-structure of the power system as well as to quantify its vulnerability to system stress. However, these indices quantify power system vulnerability using a single number, which may not convey meaningful information to an operator in a power systems control center who is trying to save the system from collapsing. Instead, it might be more beneficial to provide actionable information, such as, the amount of power transfer that must be reduced through a transmission interconnection to reduce system stress.

In [3], graph based network flow algorithms were used to detect the cut-set of minimum size between a source-sink pair of the power system. An algorithm for detecting coherent cut-sets was presented in [10]. In a coherent cut-set, power flows in all branches of the cut-set are unidirectional. The research done in this paper is aimed at finding if a contingency will create a saturated cut-set in the power network, independent of the directions in which power flows through different edges of the cut-set. The complexity of the problem lies in the fact that a power system asset can be associated with innumerable cut-sets. Therefore, the research question being explored here is: *how to analyze the power transfer capability across all cut-sets associated with an asset, and quickly screen out the cut-set that will become saturated by the largest margin as a consequence of the loss of the asset?* This paper employs exhaustive graph search algorithms to precisely answer this research question.

## II. Graph Based Network Analysis

The power system is represented by a graph $\mathcal{G}(V, E)$, such that the buses are contained in the vertex set, $V$, and the transmission lines and transformers are contained in the edge set, $E$. The generators and loads are the *sources* and *sinks*, respectively. The focus here is on transmission assets (transmission lines and transformers) and every asset has an associated capacity (referred to as the *asset rating*). Every edge $e_l \in E$ is associated with a weight $r_l$, where $r_l$ denotes maximum power that can be transferred through edge $e_l$. For simplicity, power factor is assumed to be unity (reactive power flows are not considered), and losses are considered negligible. The flow through any edge of the graph is denoted by $f_l$. The notations can be understood with reference to Figure 1. For example, edge $e_5$ joining buses 4 and 5 has a flow of 15 units from bus 4 towards bus 5 (i.e., $f_5 = 15$), and the maximum power transfer capacity of the edge is 250 units (i.e., $r_5 = 250$).

### A. Introduction to saturated cut-sets

A cut-set is defined as the set containing minimum number of edges which when removed splits the network into two disjoint islands. Any cut-set which transfers more power from one area to another than is permissible by the maximum power transfer capability of the cut-set is called a *saturated* cut-set. Let edges $e_1, e_2, \ldots, e_z$ belong to cut-set $K$. If the flows through the different edges of cut-set $K$ are $f_1, f_2, \ldots f_z$, and the ratings of those edges are $r_1, r_2, \ldots r_z$, cut-set $K$ is called a saturated cut-set if the following equation holds true:

$$\sum_{i=1}^{z} f_i > \sum_{i=1}^{z} r_i \ , \forall e_i \in K \qquad (1)$$

where $\sum_{i=1}^{z} f_i = F_K$ is the actual power flow occurring through cut-set $K$ and $\sum_{i=1}^{z} r_i = R_K$ is the maximum power that can be transferred across cut-set $K$. The objective of this research is to find if a contingency creates a saturated cut-set in the power system. If the outage of any edge $e_i \in K$ exhausts the power transfer capability of cut-set $K$, then the loss of edge $e_i$ is said to saturate cut-set $K$ by a margin of $R_K - F_K$.

This research has been funded by the PSERC projects S-74, and S-87.

The concept of saturated cut-sets is explained with the help of Figure 1. The cut-set $K^1$ in Figure 1 contains edges $e_4$, $e_6$, and $e_7$; i.e., $K^1 = \{e_4, e_6, e_7\}$. Total power transferred across this cut-set is $F_{K^1} = f_4 + f_6 + f_7 = 360$ MW. The total power transfer capacity across this cut-set is $R_{K^1} = r_4 + r_6 + r_7 = 580$ MW. It is easy to observe that the cut-set $K^1$ is unsaturated as $F_{K^1} < R_{K^1}$. However, the loss of edge 3-4 would saturate cut-set $K^1$. This is because with the outage of edge 3-4, the power that must be transferred from Area 1 to Area 2 (to satisfy the total load with total generation) is still 360 MW (i.e., $F_{K^1} = 360$), but the total power transfer capability of cut-set $K^1$ reduces to 330 MW (as now $R_{K^1} = r_6 + r_7$). Consequently, outage of edge 3-4 saturates cut-set $K^1$ by a margin of 30 MW ($R_{K^1} - F_{K^1} = 330 - 360 = -30$ MW).

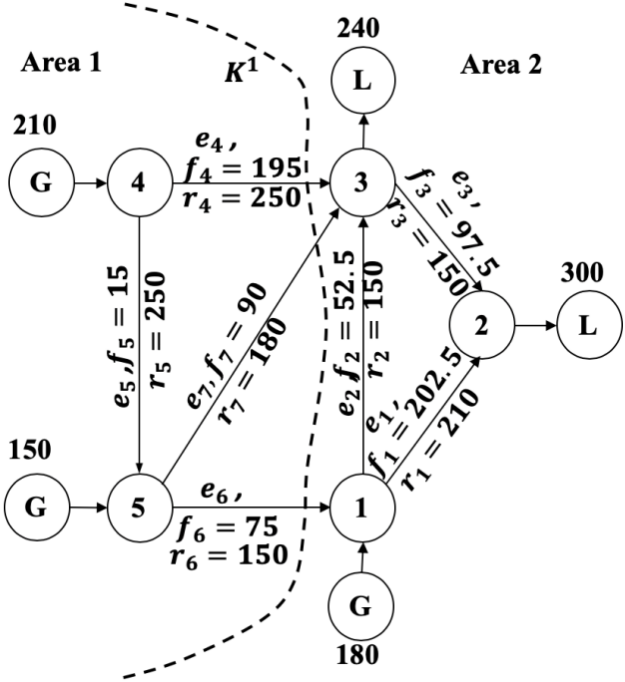

Figure 1: Effect of a contingency on a cut-set of the power network.

It must be noted here that a single edge, e.g., edge 3-4 in Figure 1, can be associated with multiple cut-sets, such as, $K^2 = \{3\text{-}4, 4\text{-}5\}$, $K^3 = \{3\text{-}4, 3\text{-}5, 1\text{-}3, 1\text{-}2\}$, and $K^4 = \{3\text{-}4, 3\text{-}5, 1\text{-}3, 2\text{-}3\}$. This implies that to assess the impact of the loss of any asset on any cut-set of the power system (to check whether it has become saturated or not), we must examine the power transfer capability of *all cut-sets associated with that asset*. For a big system containing thousands of buses, a single asset could be associated with hundreds of cut-sets. Therefore, quantifying the impact of an outage on any cut-set of the power network is a computationally intensive task.

It must also be pointed out here that when an outage creates a saturated cut-set, a DC or AC power flow solution is not even necessary to conclude that there will be post-contingency overloads. On the contrary, if an outage does not create any saturated cut-set, it may or may not result in post-contingency overloads, depending upon an actual DC or AC power flow solution (which considers the network admittances). Therefore, the proposed approach is not tailored towards identifying all possible overloads; however, when it detects a violation (i.e., saturation of a cut-set), it is guaranteed that the violation will manifest. Furthermore, if done quickly, an operator using the proposed approach will be better prepared to address the identified violations, long before the set of all possible violations are detected by a more accurate analysis. Finally, the goal here is to assess the impact of an outage on any cut-set of the power system, which is different from the information that a conventional contingency analysis scheme provides.

### B. Graph based network-flows

The objective here is to quickly find if any contingency would create a saturated cut-set in the power system. In this sub-section, the following research question is explored: *can a relaxed graph based network flow scheme, which does not consider the admittances of the system, detect saturated cut-sets*? The relaxed graph based network flow scheme is based on the following principle: *satisfy the total power demand of the loads (sinks) with the total power generated by the generators (sources), without overloading any transmission asset*. This can be easily achieved by randomly selecting a source-sink pair in the graph and subsequently satisfying the power demand of the sink using the power generation of the source via any set of paths from the source to the sink. Therefore, graph based flows respects the boundary conditions of the system (because the total generation equals the total demand) but does not consider the admittances of the network. As such, it can be intuitively realized that there can be multiple valid graph based network flow solutions. *Au contraire*, a DC or an AC power flow solution is always unique because it considers the admittances of the network. We now state the following lemma.

**Lemma:** *Any valid graph based network flow solution will result in the same power transfer across any cut-set of the power system.*

Let the network graph $\mathcal{G}(V, E)$ be split into any two clusters $C_1$ and $C_2$ such that $C_1 \cup C_2 = V$ and $C_1 \cap C_2 = \emptyset$. If $P_G^1(P_G^2)$ and $P_L^1(P_L^2)$ be the total generation and total demand in $C_1(C_2)$, then the net power generation in $C_1$ is given by $\Delta P_1 = P_G^1 - P_L^1$, while the net generation in $C_2$ is given by $\Delta P_2 = P_G^2 - P_L^2$. Cut-set $K$ between clusters $C_1$ and $C_2$ would include only those edges whose one end belongs to $C_1$ and the other end belongs to $C_2$; let the number of edges in cut-set $K$ be $z$. Also, let $f_1^A$, $f_2^A$,..., $f_z^A$ denote network flows through different edges of cut-set $K$ for a valid graph based flow solution $A$, and $f_1^B$, $f_2^B$,..., $f_z^B$ denote the network flows through the same edges for another valid graph based flow solution $B$. Then, by the law of conservation of energy, total power transfer across cut-set $K$ for each of the flow solutions $A$ and $B$ is given by,

$$\sum_{i=1}^{z} f_i^A = \sum_{i=1}^{z} f_i^B = \Delta P_1 = -\Delta P_2 \qquad (2)$$

In accordance with (2), any valid graph based network flow solution will maintain the same power transfer across any cut-set of the system. This is also numerically explained with the help of the system in Figure 1. The flows depicted in Figure 1 are obtained using the DC power flow solution computed based on the power transfer distribution factors (PTDFs) obtained using the branch admittances given in Table I.

**Table I: Branch admittances of the five-bus system**

| Branch | Admittance | Branch | Admittance |
|---|---|---|---|
| 1-2 | $2x$ | 3-5 | $x$ |
| 1-3 | $x$ | 4-5 | $2x$ |
| 1-5 | $2x$ | 3-4 | $2x$ |
| 2-3 | $2x$ | | |

*Branch admittances are represented in terms of a variable $x$

Now, for the same system, multiple valid graph based network flow solutions can be obtained by combining different source-sink pairs. Figure 2 and Figure 3 depict two valid graph based network flow solutions.

*Creating Figure 2:* The steps involved in creating the *flow graph* of Figure 2 are as follows: (i) use 150 MW of power injection at the source vertex 5 to satisfy 150 MW of power demand at the sink vertex 3, along path 5-3, (ii) use 90 MW of power injection at the source vertex 4 to satisfy the remaining 90 MW of power demand at the sink vertex 3, via path 4-3, (iii) use the remaining 120 MW of power injection at the source vertex 4 to satisfy 120 MW of power demand at the sink vertex 2 via path 4-3-2, and finally (iv) use 180 MW of power generation at the source vertex 1 to satisfy 180 MW of power demand at the sink vertex 2, via path 1-2.

*Creating Figure 3:* The steps involved in creating the *flow graph* of Figure 3 are enumerated as follows: (i) use 150 MW of power injection of the source vertex 5 to satisfy 150 MW of power demand at the sink vertex 2 along path 5-1-2, (ii) use 60 MW of power injection of the source vertex 1 to satisfy 60 MW of power demand at the sink vertex 2 along path 1-2, (iii) use 90 MW of power injection at the source vertex 1 to satisfy the remaining 90 MW of power demand at the sink vertex 2 via path 1-3-2, (iv) use 30 MW of remaining power injection from the source vertex 1 to satisfy 30 MW of power demand at the sink vertex 3, and finally (v) use 210 MW of generation from the source vertex 4 to satisfy the remaining 210 MW of demand at the sink vertex 3 via path 4-3.

Now, it is observed that even though the branch flows in the three flow graphs of Figure 1, Figure 2, and Figure 3 are different, the net power transfer across any cut-set of the network is the same. For example, Table II depicts the total power transfer across cut-set $K^1$={4-3,5-3,5-1} for the three flow graphs of Figure 1, Figure 2, and Figure 3, respectively. Although the individual flows through different edges of cut-set $K^1$ are different, the total power transfer across cut-set $K^1$ is same (= 360 MW) for each of the three flow graphs.

Based on these observations, we conclude that by not considering the admittances of the network it is possible to achieve multiple valid graph based flow solutions, all of which maintain the same power transfer across any cut-set of the network. As the net power transfer across any cut-set for each of the flow graphs is the same, any valid flow graph can be used to assess the impact of an outage on a cut-set of the network. This ability to ignore the network admittances but still obtain meaningful information facilitates the enhancement in computational speed of the proposed approach. Lastly, while selecting a path from a source to a sink in the different iterations of the graph based network flow scheme, *the shortest path* from source to sink is selected for additional computational benefits. In this research, the shortest path has been selected using the breadth first search (BFS) graph traversal algorithm [11].

### C. Identification of saturated cut-sets

To investigate if the loss of any asset saturates a cut-set, **Algorithm I** is proposed. It screens out the cut-set $K_{crit}$ which is saturated by the *largest margin* due to the outage of edge $e_l$. In addition, the algorithm also finds the margin, $T_m^l$, by which the cut-set becomes saturated. **Algorithm I** is named "feasibility test (FT)" because it investigates the feasibility of an outage and assesses its impact on any cut-set of the power system. Although the edge $e_l$ can be associated with multiple cut-sets, **Algorithm I** is able to screen out the cut-set that becomes saturated by the largest margin, because that cut-set is the first one to get saturated in Step (ii) of the algorithm.

Table II: Power transfer across cut-set $K^1$ for three different network flow solutions

| Edges in $K^1$ | Figure 1 Flow (MW) | Figure 2 Flow (MW) | Figure 3 Flow (MW) |
|---|---|---|---|
| 4-3 | 195 | 210 | 210 |
| 5-3 | 90 | 150 | 0 |
| 5-1 | 75 | 0 | 150 |
| Power transfer | 360 | 360 | 360 |

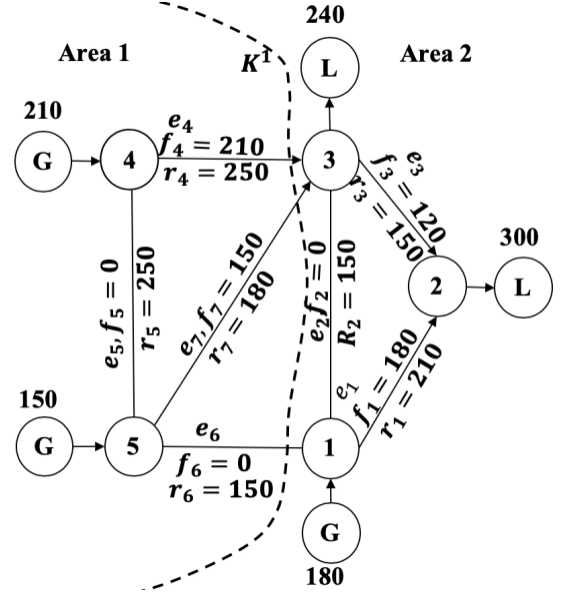

Figure 2: A flow graph based upon graph based network flow solution.

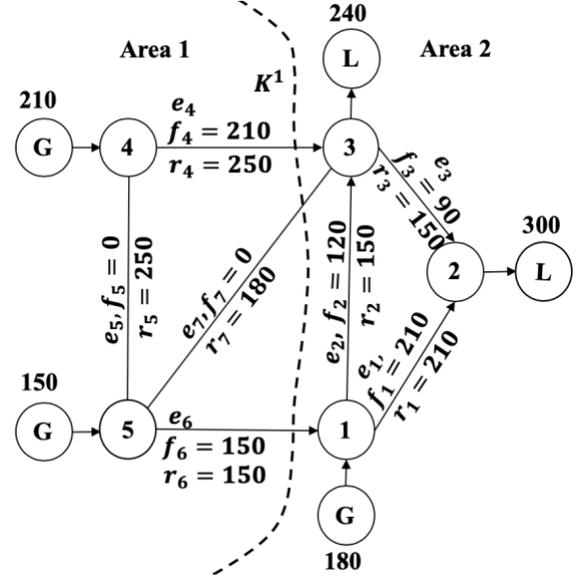

Figure 3: Another flow graph based upon graph based network flow solution.

If **Algorithm I** is applied to any of the three flow graphs of Figure 1, Figure 2, or Figure 3 to investigate the outage of edge $e_4$, the following conclusion will be reached independent of which flow graph is used for the analysis: *the outage of edge 3-4 ($e_4$) would maximally saturate cut-set $K^1$ by a margin of 30 MW*, i.e. $K_{crit} = K^1$ and $T_m^4 = -30$ MW. This is explained with help of Figure 4. Figure 4 depicts the power transfer across all four cut-sets associated with edge 3-4 for the flow graph of Figure 1. Consider cut-set $K^3$ in Figure 4(c). The total power transfer across $K^3$ is 540 MW ($F_{K^3} = f_1 + f_2 + f_4 + f_7$) and the maximum power that can be transferred across $K^3$ after the outage of 3-4 is also 540 MW ($R_{K^3} = r_1 + r_2 + r_7$). As such, outage of 3-4 saturates $K^3$ by 0 MW ($R_{K^3} = F_{K^3}$). Similarly, it can be observed from Figure 4 that the outage of 3-4 creates a negative margin of 30 MW in cut-set $K^1$, positive margin of 40 MW in $K^2$, and a positive margin of 240 MW in $K^4$. Therefore, it is validated using Figure 4 that **Algorithm I** correctly identifies the cut-set which gets saturated by the largest (negative) margin. Identical results were obtained when **Algorithm I** was applied to the flow graphs of Figure 2 and Figure 3. Lastly, it must also be pointed out that the margin computed by **Algorithm I** is indicative of the minimum amount of power transfer that must be reduced across cut-set $K_{crit}$ to alleviate its saturation due to the contingency.

**Algorithm I**: Feasibility test (FT)
i. Let $f_l^{FT}$ units of power flow through edge $e_l$ from vertex $v_l^F$ to vertex $v_l^T$. Remove edge $e_l$ from the original graph of the network and initialize a variable $TC_l$ to zero.
ii. Search the graph to obtain the shortest unsaturated path $P$ from $v_l^F$ to $v_l^T$ using breadth first search (BFS) [11]. Path $P$ is considered unsaturated if it has capacity to reroute additional flow.
iii. Find the maximum extra flow $C_P$, that can be re-routed through path $P$. Update $TC_l \coloneqq TC_l + C_P$ and inject $C_P$ units of flow along path $P$. Note that this step saturates path $P$ of the graph.
iv. Repeat steps (ii) and (iii) until there exists no unsaturated path from $v_l^F$ to $v_l^T$ or $f_l^{FT}$ units of flow have been re-routed.
v. Compute the transfer margin $T_m^l$ for the outage of $e_l$ as: $T_m^l = TC_l - f_l^{FT}$. If $T_m^l$ is negative, loss of $e_l$ creates a saturated cut-set.
vi. To identify $K_{crit}$, traverse the saturated graph from $v_l^F$ to $v_l^T$. All the vertices that *can be reached* from $v_l^F$ without traversing a saturated edge are grouped into cluster $C_1$. Similarly, the vertices that *cannot be reached* from $v_l^F$ without traversing a saturated edge are grouped into cluster $C_2$. The cut-set $K_{crit}$ contains the set of edges whose one end belongs to $C_1$ and the other end belongs to $C_2$.

## III. RESULTS AND DISCUSSION

### A. Identification of saturated cut-sets in IEEE 39-bus system

The system data for the IEEE 39-bus test system is obtained from MATPOWER [12]. When every transmission asset was investigated by the graph based feasibility test (FT), four saturated cut-sets were identified which are depicted by dotted lines in Figure 5. The detailed information obtained from the graph based analysis is summarized below:
i. Outage of 11-10 saturates cut-set $K_{crit}^1=\{11\text{-}10,13\text{-}10\}$ by 61 MW. Similarly, outage of 13-10 saturates the same cut-set $K_{crit}^1$ by the same margin, because both the lines 11-10 and 13-10 have the same rating of 600 MVA.
ii. Outage of 6-11 saturates cut-set $K_{crit}^2=\{6\text{-}11,14\text{-}13\}$ by 52 MW. However, outage of 14-13 saturates the same cut-set $K_{crit}^2$ by 172 MW. This is because branches 6-11 and 14-13 have ratings of 480 MVA and 600 MVA, respectively.
iii. Outage of 21-22 saturates cut-set $K_{crit}^3=\{21\text{-}22,24\text{-}23\}$ by 393 MW, and the outage of 24-23 saturates $K_{crit}^3$ by 93 MW. This is because branch 21-22 has a rating of 900 MVA, while branch 24-23 has a rating of 600 MVA.
iv. Outage of 16-21 saturates cut-set $K_{crit}^4=\{16\text{-}21,24\text{-}23\}$ by 119 MW. Similarly, outage of 24-23 has the same effect on $K_{crit}^4$, because both lines have same ratings.

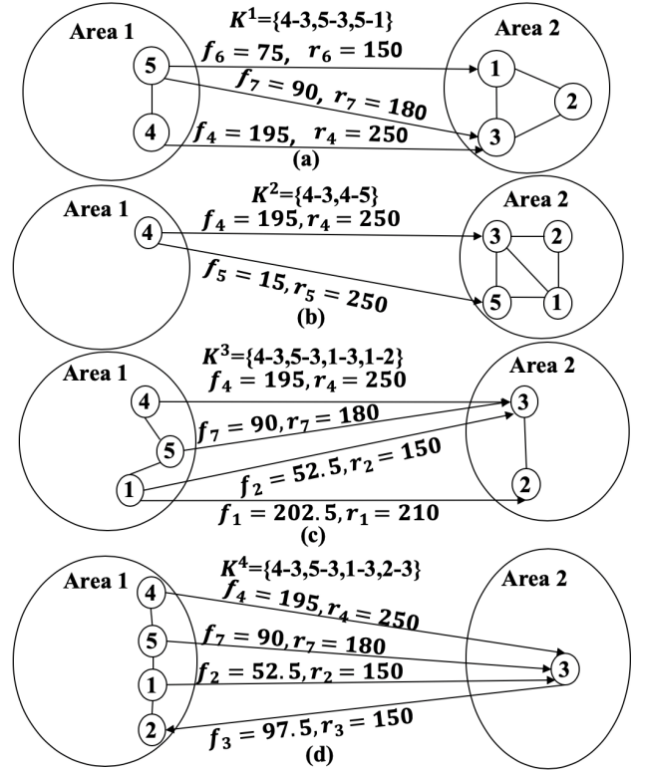

Figure 4: Effect of the outage of edge 3-4 on (a) $K^1$, (b) $K^2$, (c) $K^3$, and (d) $K^4$ of the flow graph of Figure 1.

This ability to assess the impact of an outage on any cut-set of the power network and screen out the cut-set that becomes saturated by the largest margin facilitates creation of an enhanced power system connectivity monitoring scheme and visualization platform for better situational awareness in a power systems control room. Also note that this insight is very different from what a conventional contingency analysis scheme may provide. Moreover, the proposed analysis not only identifies the saturated cut-set, but also indicates the minimum amount of power transfer that must be reduced across the cut-set to alleviate its saturation. For the case-study shown in Figure 5, for each cut-set, we increased the power generation on one side of the cut-set and reduced the generation on the other side by the calculated margins to verify that the saturation of the cut-sets were indeed alleviated. The validation was done using a DC power flow solution. For example, the outage of branch 6-11 saturated cut-set $K_{crit}^2$ by a margin of 52 MW. As a corrective action, the generation at bus 39 was increased by 52 MW and the generation at bus 32 was reduced by 52 MW. A DC power flow solution revealed that the saturation of the cut-set was removed due to the change in generation dispatch. However, this action resulted in an overload of 47 MW on line 4-14. An intelligent technique for initiating a corrective action that does not result in subsequent network violations would be investigated in the future.

### B. Computation time of graph based feasibility test (FT)

The objective here is to make the graph based network analysis fast enough for real-time power systems operations. As such, the computational time of the proposed graph based network analysis is compared with DC contingency analysis. Figure 6 compares the computational time of graph based FT and DC contingency analysis for a N-1 analysis. A N-1 analysis examines the outage of every single transmission asset in the power system. The computational time along the ordinate of Figure 6 is presented in the log-scale (base of 10), while the abscissa denotes the test systems of different sizes. It is clear from Figure 6 that as the system size increases the computation time of DC contingency analysis increases at a much faster rate in comparison to the graph based FT. Notably, for test systems containing approximately 2,000-3,000 buses, the computation time of N-1 DC contingency analysis is of the order of $10^3$ s, while that of the graph based FT is only $10^1$ s (a two orders of magnitude improvement in speed). The graph based N-1 FT and DC contingency analysis were done on the same computer with 6 GB RAM and 2.3 GHz Core i5 processor.

It can be argued that using the line outage distribution factors (LODFs), which can be stored offline, DC contingency analysis can be done even faster [13]. However, the scope of this research relates to situations when multiple outages occur in rapid succession, during which, the LODFs computed offline cannot be used for the analysis. As such, the fast computation time of the proposed graph based network analysis shows good prospects when enhanced situational awareness is required in a limited timeframe (e.g. when successive outages occur during extreme event scenarios).

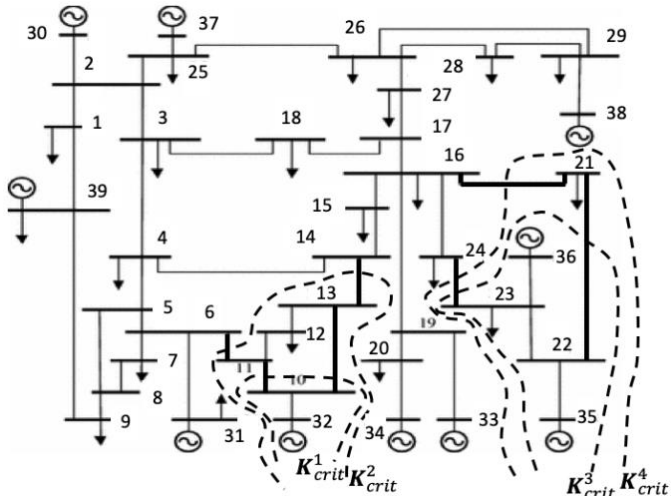

Figure 5: Identification of critical cut-sets in the IEEE 39-bus system.

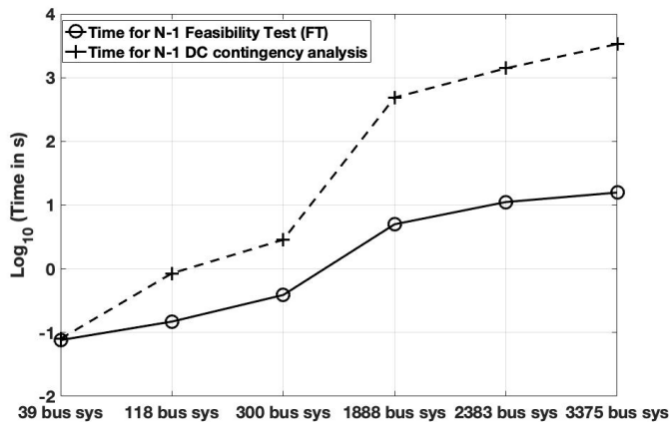

Figure 6: Comparison of computation time of feasibility test (FT) and DC contingency analysis.

## IV. CONCLUSIONS AND FUTURE WORK

This paper proposed a new graph theoretic network analysis scheme to detect if a contingency will create any saturated interconnection (or cut-set) in the power network. The fast computation of the proposed algorithm makes it suitable for real-time power systems operations. The proposed graph based network analysis tool can assess the impact of an outage on different cut-sets of the power system and then screen out the cut-set which becomes saturated by the largest margin. The computed margin indicates the minimum amount by which the power transfer across the critical interconnection must be reduced. The research presented here lays the foundation for the following questions that can be explored in the future:

i. During a multi-failure scenario, after a branch outage has occurred, how can the flow graph be updated quickly, without having to recreate it from scratch?
ii. It is unlikely that an outage in one part of the network will affect a portion of the network which is very far away. Is it possible to develop an intelligent shortlisting scheme to determine which assets must be examined by FT during successive failures?
iii. Moreover, once the saturated cut-sets are identified by the proposed graph based network analysis what should be the best strategy to initiate a corrective action that would alleviate the saturation of the cut-set without creating other network violations (e.g. overloads)?
iv. Since there might be multiple saturated cut-sets at a given point in time how to ensure that a single corrective action would alleviate the saturation of all cut-sets in the system?

The above-mentioned research questions will be explored using a 20,000+ bus model of the Western Interconnection (WI).

REFERENCES

[1] FERC and NERC staff "Arizona-Southern California outages on September 8, 2011, causes and recommendations," *Federal Energy Regulatory Commission* and *North American Electric Reliability Corporation*, Apr. 2012. [Online] Available: https://www.ferc.gov/legal/staff-reports/04-27-2012-ferc-nerc-report.pdf
[2] FERC Staff, "The Con Edison power failure of July 13 and 14, 1977," *U.S. Department of Energy Federal Energy Regulatory Commission*, pp. 2-3, 19-20 Jun. 1978.
[3] T. Werho, V. Vittal, S. Kolluri, and S. M. Wong, "Power system connectivity monitoring using a graph theory network flow algorithm," *IEEE Trans. Power Syst.*, vol. 31, no. 6, pp. 4945-4952, Nov. 2016.
[4] J. Beyza, E. Garcia-Paricio, and J. M. Yusta, "Applying complex network theory to the vulnerability assessment of interdependent energy infrastructures," *Energies*, vol. 12, no. 3, Jan. 2019.
[5] H. Bai, and S. Miao, "Hybrid flow betweenness approach for identification of vulnerable line in power system," *IET Gener. Transm. Distrib.*, vol. 9, no. 12, pp. 1324-1331, Jan. 2015.
[6] E. Bompard, E. Pons, and D. Wu, "Extended topological metrics for the analysis of power grid vulnerability," *IEEE Syst. J.*, vol. 6, no. 3, pp. 481-487, Sep. 2012.
[7] Y. Zhu, J. Yan, Y. Sun, and H. He, "Revealing cascading failure vulnerability in power grids using risk-graph," *IEEE Trans. Parallel Distrib. Syst.*, vol. 25, no. 12, pp. 3274–3284, Dec. 2014.
[8] J. Beyza, J. Yusta M, G. Correa J, and H. Ruiz F, "Vulnerability assessment of a large electrical grid by new graph theory approach," *IEEE Latin America Trans.*, vol. 16, no. 2, pp. 527-535, Feb. 2018.
[9] J. Fang, C. Su, Z. Chen, H. Sun, and P. Lund, "Power system structural vulnerability assessment based on an improved maximum flow approach," *IEEE Trans. Smart Grid*, vol. 9, no. 2, pp. 777-785, Mar. 2018.
[10] A. Beiranvand, and P. Cuffe, "A topological sorting approach to identify coherent cut-sets within power grids," accepted in *IEEE Transactions on Power Systems*.
[11] D. Angel, "A breadth first search approach for minimum vertex cover of grid graphs," in *Proc. IEEE 9th Int. Conf. Intelligent Syst. Control (ISCO)*, Coimbatore, India, pp. 1-4, 9-10 Jan. 2015.
[12] R. D. Zimmerman, "MATPOWER 4.0b4 User's Manual," [Online]. Available: www.pserc.cornell.edu/matpower
[13] T. Guler, G. Gross and M. Liu, "Generalized line outage distribution factors," *IEEE Trans. Power Syst.*, vol. 22, no. 2, pp. 879-881, May 2007.